\documentclass[10pt,twocolumn]{article}

\usepackage[
  a4paper,
  top=2.0cm,
  bottom=2.0cm,
  left=1.7cm,
  right=1.7cm,
  columnsep=0.7cm
]{geometry}
\usepackage[T1]{fontenc}
\usepackage{lmodern}
\usepackage{graphicx}
\usepackage{amsmath,amssymb}
\usepackage[numbers,square,sort&compress]{natbib}
\usepackage{setspace}
\usepackage{microtype}
\usepackage{xurl}
\usepackage{hyperref}
\usepackage{authblk}
\usepackage{orcidlink}

\usepackage{caption}
\DeclareCaptionLabelSeparator{edbar}{\enspace\textbar\enspace}

\hypersetup{
  colorlinks=true,
  citecolor=blue,
  linkcolor=blue,
  urlcolor=blue,
  pdftitle={Evidence for orbital eccentricity supports a hierarchical origin for GW231123}
}
\setcitestyle{numbers,square,comma}
\title{Evidence for orbital eccentricity supports a hierarchical origin for GW231123}

\author[1]{Baoxiang Wang\,\orcidlink{0009-0003-3998-4609}}

\author[1,*]{Tao Yang\,\orcidlink{0000-0002-2161-0495}}

\affil[1]{School of Physics and Technology, Wuhan University, Wuhan 430072, China}

\author[2]{Zhoujian Cao\,\orcidlink{0000-0002-1932-7295}}

\author[2]{Bin Hu\,\orcidlink{0000-0001-5093-8118}}

\affil[2]{School of Physics and Astronomy, Beijing Normal University, Beijing 100875, China}

\author[3]{Xikai Shan\,\orcidlink{0000-0003-3201-061X}}
\affil[3]{Max Planck Institute for Gravitational Physics (Albert Einstein Institute), Am Mühlenberg 1, Potsdam 14476, Germany}

\author[4,5]{Rong-Gen Cai\,\orcidlink{0000-0002-3539-7103}}
\affil[4]{Zhejiang Key Laboratory of Extreme Universe, Ningbo University, Ningbo 315211, China}
\affil[5]{Institute of Fundamental Physics and Quantum Technology, Ningbo University, Ningbo 315211, China}

\date{}

\begin{document}

\twocolumn[
\begin{@twocolumnfalse}
\vspace{-2em}
\maketitle
\vspace{-3em}
\begin{abstract}
\normalsize

GW231123 is the most massive binary black hole merger observed to date and one of the most intriguing hierarchical merger candidates in the pair-instability mass gap. Its exceptionally large inferred component spins and remarkably short in-band duration make the source highly sensitive to waveform physics near merger. Here we show that GW231123 is better described as an eccentric binary black-hole merger. We analyse data over 11--448 Hz using the eccentric, spin-precessing effective-one-body waveform model TEOBResumS--Dal\'i, accounting for detector calibration uncertainty. We measure $e_{5.5\,\mathrm{Hz}}=0.34^{+0.12}_{-0.23}$
at 90\% credibility and obtain a Bayes factor of 10.2 in favour of the eccentric model over its quasi-circular counterpart.
GW231123-like injection--recovery studies further show that analyses starting at 20 Hz, as in previous studies, can obscure  this eccentricity.
Including eccentricity substantially changes the spin inference, weakening the preference for a near-extremal secondary spin and lowering its posterior median to $a_2\simeq0.7$, close to the characteristic spin expected for a black-hole
merger remnant.
These results provide a coherent dynamical interpretation of GW231123, in which residual eccentricity, a mass-gap black hole, and a
remnant-like spin arise naturally from a hierarchical merger channel
in a dense environment.

\end{abstract}
\vspace{1em}
\end{@twocolumnfalse}
]

\begingroup
\renewcommand{\thefootnote}{\fnsymbol{footnote}}
\footnotetext[1]{Corresponding author:
\href{mailto:yangtao@whu.edu.cn}{yangtao@whu.edu.cn}}
\endgroup

On 23 November 2023, the LIGO Hanford and Livingston
observatories detected the gravitational event GW231123, the most massive binary
black-hole (BBH) merger reported to date
\cite{LIGOScientific:2025rsn}.
Under the quasi-circular interpretation, the source-frame component
masses are $137^{+23}_{-18}\,M_\odot$ and
$101^{+22}_{-50}\,M_\odot$, with the secondary overlapping the pair-instability mass gap and the primary overlapping its upper boundary. The inferred component-spin magnitudes are also large, $0.9^{+0.10}_{-0.19}$ and $0.80^{+0.20}_{-0.52}$ at 90\% credibility \cite{LIGOScientific:2025rsn}. Together with its remarkably short in-band duration, these properties make GW231123 both an intriguing candidate for
hierarchical assembly and a particularly challenging event to interpret.

Black holes in the pair-instability mass gap are difficult to produce through ordinary stellar collapse
\cite{Heger:2001cd,Woosley:2016hmi,Tong:2025wpz}.
Alternative formation pathways include hierarchical mergers in dense stellar systems and active galactic nucleus discs, where merger remnants can be retained and participate in subsequent mergers
\cite{Rodriguez:2017pec,Gerosa:2021mno,Bartos:2016dgn,Samsing:2020tda}. For comparable-mass binaries with small progenitor spins, the merger remnant has a characteristic dimensionless spin near $a\simeq0.7$, although the remnant-spin distribution depends on the progenitor masses and spins~\cite{Gerosa:2021mno}.
The combination of mass-gap components and the very large spins inferred for GW231123 therefore raises an important question: whether these properties reflect the source itself or, at least in part, the waveform assumptions used to infer them.

This question takes on added significance for GW231123 because only a few measurable waveform cycles are observed before merger. Its inferred source properties are consequently sensitive to waveform modelling in the late inspiral and merger regime
\cite{LIGOScientific:2025rsn,Jan:2025zcm, Chandra:2026voe}.
Orbital eccentricity is especially relevant in this context, considering the possibility that GW231123 is a hierarchical merger. Gravitational radiation tends to circularize long-lived isolated binaries, whereas gravitational-wave captures, binary-single encounters and secular interactions in dense environments can produce binaries that retain measurable eccentricity in the ground-based detector band
\cite{Peters:1964zz,Samsing:2017xmd,Antonini:2013tea,Chattopadhyay:2023pil}.
Eccentricity modifies the phase and amplitude structure of the waveform and can be difficult to distinguish from spin precession when only a few merger-dominated cycles are observed
\cite{Romero-Shaw:2022fbf,Tibrewal:2026jci}. When non-circular structure is omitted or only weakly constrained,
part of its waveform imprint may be accommodated by the spin degrees of freedom, potentially biasing the inferred spins \cite{Divyajyoti:2025cwq,
Chandra:2026voe}.
This raises the question of whether its extreme inferred spins depend in part on the assumed orbital dynamics.

Here we analyse GW231123 over 11--448\,Hz using the eccentric, spin-precessing effective-one-body waveform model TEOBResumS--Dal\'i, including higher-order modes and detector calibration uncertainty. We initialize the waveform and define eccentricity at 5.5\,Hz,
fully utilizing the available low-frequency information. We use the publicly released glitch-subtracted Hanford strain down to 11 Hz and adopt a conservative 20 Hz low-frequency cutoff for Livingston, as detailed in Methods. Within the same precessing waveform framework, we compare eccentric and quasi-circular hypotheses and find evidence favouring an eccentric interpretation of GW231123. Including eccentricity selectively weakens the preference for a near extremal secondary spin and lowers its posterior median to $a_2\simeq0.7$, close to the characteristic spin expected for a black hole merger remnant.
We further perform GW231123-like injection--recovery analyses over 11--448\,Hz and 20--448\,Hz, obtaining eccentricity constraints consistent with the corresponding real data results in both frequency bands. Our results provide an eccentric interpretation of GW231123 and link the event to a possible hierarchical-merger origin.

\section*{Results}

\subsection*{Evidence for orbital eccentricity}

We compare eccentric and quasi-circular hypotheses using H1 data over 11--448 Hz and L1 data over 20--448 Hz, with identical calibration treatment and priors for all shared parameters. Both models include spin
precession and higher-order modes, while the eccentric model
additionally allows the orbital eccentricity and mean anomaly to
vary.
We obtain a log Bayes factor between the eccentric and
quasi-circular hypotheses of
$\ln \mathcal{B}_{\rm ecc/qc}
=\ln Z_{\rm ecc}-\ln Z_{\rm qc}=2.32$,
corresponding to a Bayes factor of $10.2$ in favour of the eccentric model.

\begin{figure}[t]
  \centering
  \includegraphics[width=\columnwidth]{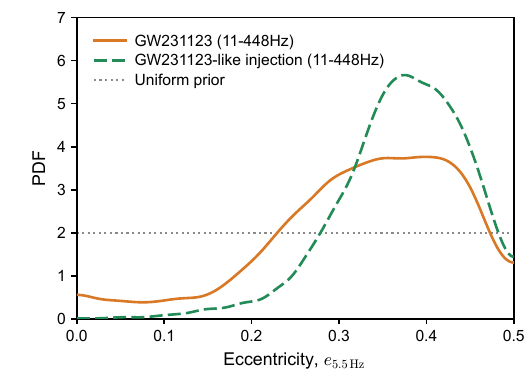}
  \caption{
  \textbf{Eccentricity of GW231123 in the 11--448 Hz analysis.}
  Marginalized $e_{5.5\,\mathrm{Hz}}$ posteriors for GW231123
and the GW231123-like eccentric injection recovery, both
obtained in the 11--448\,Hz analysis.
The dotted horizontal line shows the common uniform prior density.
GW231123 gives
$e_{5.5\,\mathrm{Hz}}=0.34^{+0.12}_{-0.23}$,
while the injection recovery gives
$e_{5.5\,\mathrm{Hz}}=0.37^{+0.09}_{-0.14}$.}
  \label{fig:e55}
\end{figure}

The marginalized posterior shown in Fig.~\ref{fig:e55} gives
$e_{5.5\,\mathrm{Hz}}=0.34^{+0.12}_{-0.23}$.
Unless otherwise stated, we report all measurements as the median and 90\% credible interval.
Here $e_{5.5\,\mathrm{Hz}}$ denotes the eccentricity at the
5.5 Hz waveform reference frequency.

\subsection*{Eccentricity inference at 10\,Hz from the 20--448\,Hz band}

To facilitate comparison with previous studies, we repeat the eccentricity analysis using data from both detectors over 20--448 Hz, with eccentricity defined at 10 Hz. We obtain
$e_{10\,\mathrm{Hz}}=0.12^{+0.27}_{-0.11}$.
Our 20--448\,Hz result is consistent with previous analyses of
GW231123 over the same frequency range, with no
compelling evidence for nonzero eccentricity
\cite{Jan:2025zcm,Gamboa:2026ybr}.
In particular, the independent eccentric, spin-precessing
\texttt{SEOBNRv6EPHM} analysis reports
$e_{10\,\mathrm{Hz}}=0.10^{+0.18}_{-0.09}$,
in remarkably close agreement with our measurement.
As shown in Fig.~\ref{fig:e10comparison}, the GW231123 posterior
retains substantial support at small eccentricities, close to
the quasi-circular boundary.

\subsection*{Injection--recovery consistency analysis}
To test the consistency of our eccentricity inference in the two
frequency bands, we perform injection--recovery analyses of a
GW231123-like eccentric signal. 
The injection parameters are based on the joint maximum-likelihood
configuration of our 11--448 Hz eccentric analysis, for which
$e_{5.5\,\mathrm{Hz}}^{\rm inj}=0.30$,
close to the posterior median
$e_{5.5\,\mathrm{Hz}}=0.34$ inferred from the GW231123 data. 
We use this same injected signal
for the recoveries in both frequency bands.

Recovering the injection over 11--448\,Hz, with eccentricity
defined at 5.5\,Hz, gives
$e_{5.5\,\mathrm{Hz}}=0.37^{+0.09}_{-0.14}$,
in excellent agreement with the eccentricity inferred from
the GW231123 data (Fig.~\ref{fig:e55}).
The same injection yields
$\ln\mathcal{B}_{\rm ecc/qc}=2.61$,
close to the value of $2.32$ obtained from the GW231123 data,
providing an additional consistency check on the relative
model preference.
Note that previous injection studies have likewise shown that genuinely
eccentric signals need not yield a strong Bayesian preference
over quasi-circular models, particularly for short signals with
only a few observable cycles
\cite{Divyajyoti:2023rht,Romero-Shaw:2022fbf}.

\begin{figure}[t]
  \centering
  \includegraphics[width=\columnwidth]{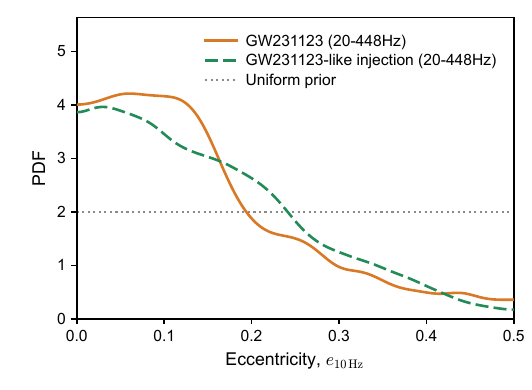}
  \caption{
  \textbf{Eccentricity posteriors in the 20--448 Hz analyses.}
  Marginalized $e_{10\,\mathrm{Hz}}$ posteriors for GW231123
  and the GW231123-like eccentric injection recovery, both
  obtained in the 20--448 Hz analysis.
  The dotted horizontal line shows the common uniform prior
  density.
  GW231123 gives
  $e_{10\,\mathrm{Hz}}=0.12^{+0.27}_{-0.11}$,
  while the injection recovery gives
  $e_{10\,\mathrm{Hz}}=0.14^{+0.24}_{-0.12}$.
  }
  \label{fig:e10comparison}
\end{figure}

Recovering the same injection over 20--448\,Hz, with eccentricity
defined at 10\,Hz, gives
$e_{10\,\mathrm{Hz}}=0.14^{+0.24}_{-0.12}$,
also consistent with the corresponding GW231123 result.
As shown in Fig.~\ref{fig:e10comparison}, both the GW231123 and
injection posteriors retain substantial support at small
eccentricities, close to the quasi-circular boundary.
The injection result shows that a signal that is eccentric at
lower frequencies can nevertheless yield substantial posterior
support close to the quasi-circular boundary when analysed only
above 20\,Hz.

Orbital eccentricity is larger at lower frequencies and decreases
approximately as an inverse power of gravitational wave frequency, making the low-frequency part of
the signal particularly valuable for revealing residual eccentricity.
Both the real data and injection--recovery analyses show this trend.
The agreement between the injection recoveries and the corresponding
GW231123 results at both reference frequencies provides a
self-consistency check and strengthens confidence in our eccentricity
inference.

\begin{figure*}[t]
  \centering
  \includegraphics[width=\textwidth]{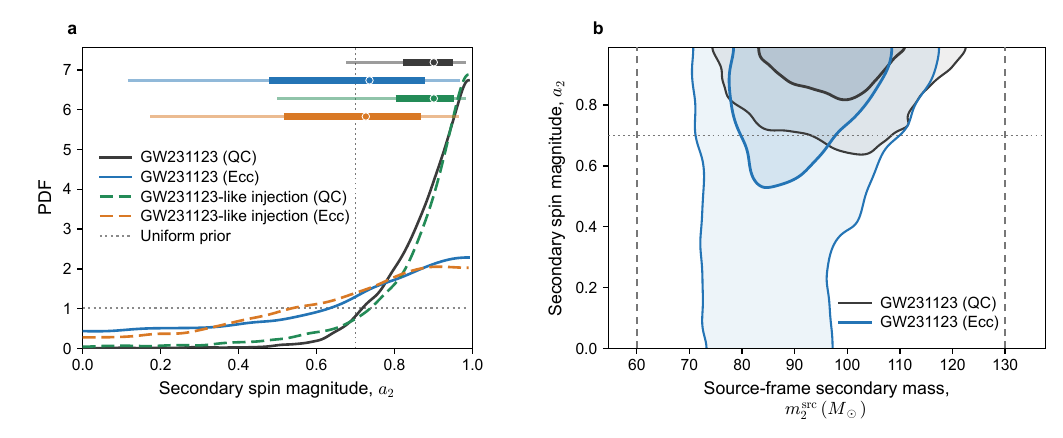}
  \caption{
\textbf{Secondary-spin inference for GW231123 and injection recovery.}
The labels QC and Ecc in parentheses denote the quasi-circular and eccentric waveform hypotheses, respectively. Both panels show results from the 11--448\,Hz analysis.
\textbf{a}, Marginalized posterior distributions of the secondary-spin
magnitude $a_2$. The black and blue solid curves show the GW231123
posteriors under the quasi-circular and eccentric hypotheses,
respectively. The green and orange dashed curves show the
quasi-circular and eccentric recoveries of the same GW231123-like
eccentric injection, respectively.
The thick and thin horizontal bars indicate the 50\% and 90\% credible intervals, respectively, and the dots mark
the posterior medians. 
The dotted horizontal line shows the uniform prior density,
while the dotted vertical line marks the reference value $a_2=0.7$.
\textbf{b}, Joint posterior distributions of the source-frame
secondary mass $m_2^{\rm src}$ and secondary-spin magnitude $a_2$
for GW231123. Black and blue solid contours correspond to the
quasi-circular and eccentric hypotheses, respectively.
The inner and outer contours enclose the 50\% and 90\%
highest-posterior-density credible regions, respectively.
The dashed vertical lines at $m_2^{\rm src}=60\,M_\odot$ and
$130\,M_\odot$ mark the boundaries of the reference pair-instability
mass-gap interval. The dotted horizontal line marks $a_2=0.7$.
}
  \label{fig:spin}
\end{figure*}

\begin{figure*}[t]
  \centering
  \includegraphics[width=\textwidth]{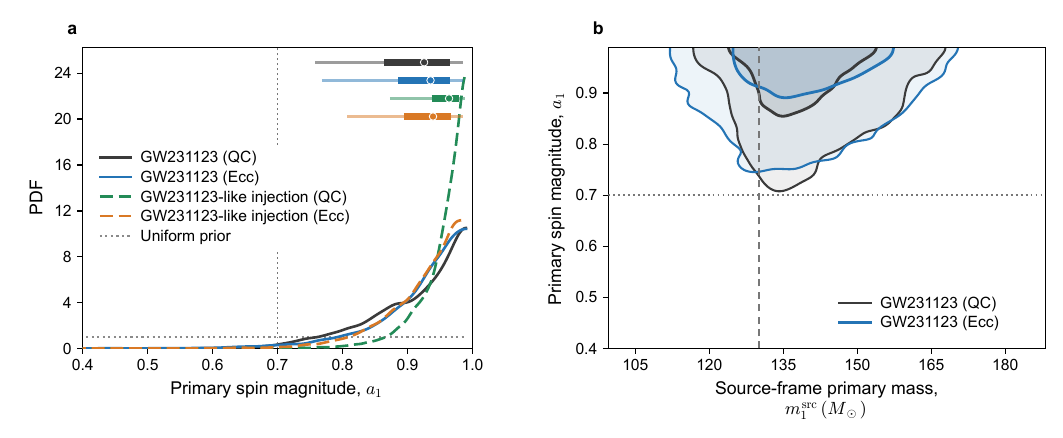}
  \caption{
\textbf{Primary-spin inference for GW231123 and injection recovery.}
The labels QC and Ecc in parentheses denote the quasi-circular and eccentric waveform hypotheses, respectively.
Both panels show results from the 11--448\,Hz analysis. For clarity, the $a_1$ axes in both panels are restricted to
$a_1\geq0.4$, as the posterior support below this value is negligible.
\textbf{a}, Marginalized posterior distributions of the primary-spin
magnitude $a_1$. The black and blue solid curves show the GW231123
posteriors under the quasi-circular and eccentric hypotheses,
respectively. The green and orange dashed curves show the
quasi-circular and eccentric recoveries of the same GW231123-like
eccentric injection, respectively.
The thick and thin horizontal bars indicate the 50\% and 90\% credible intervals, respectively, and the dots mark
the posterior medians. 
The dotted horizontal line shows the uniform prior density,
while the dotted vertical line marks the reference value $a_1=0.7$.
\textbf{b}, Joint posterior distributions of the source-frame
primary mass $m_1^{\rm src}$ and primary-spin magnitude $a_1$
for GW231123. Black and blue solid contours correspond to the
quasi-circular and eccentric hypotheses, respectively.
The inner and outer contours enclose the 50\% and 90\%
highest-posterior-density credible regions, respectively.
The dashed vertical line at $m_1^{\rm src}=130\,M_\odot$ marks
the upper boundary of the reference pair-instability mass-gap
interval. The dotted horizontal line marks $a_1=0.7$.
}

  \label{fig:primary_spin}
\end{figure*}

\subsection*{Eccentricity brings the secondary spin closer to the merger-remnant scale}

Allowing eccentricity substantially changes the inferred
secondary-spin distribution of GW231123. For the 11--448\,Hz
analysis, the quasi-circular model gives
$a_2=0.90^{+0.08}_{-0.22}$, whereas the eccentric model gives
$a_2=0.74^{+0.23}_{-0.62}$.
As shown in Fig.~\ref{fig:spin}a, the quasi-circular posterior is
strongly weighted towards large spin values, whereas allowing
eccentricity lowers the posterior median and extends the distribution
towards substantially smaller spins.

To test whether this behaviour can be reproduced for an eccentric
signal, we recover the GW231123-like eccentric injection with the
quasi-circular waveform hypothesis. The resulting secondary spin
posterior is again shifted towards large values and closely resembles
the quasi-circular posterior obtained from the GW231123 data
(Fig.~\ref{fig:spin}a). For the injection, changing from the eccentric
to the quasi-circular recovery shifts the secondary-spin posterior
median by approximately $\Delta a_2\simeq0.17$. This provides an
event-specific consistency check that omitting eccentricity can
produce the high secondary spin preference seen in the real data
quasi-circular analysis.

The response is markedly different for the primary black hole.
As shown in Fig.~\ref{fig:primary_spin}a, the primary spin posterior
remains concentrated at large values under both the quasi-circular
and eccentric hypotheses. The corresponding injection recovery shows
the same behaviour: changing the recovery from the eccentric to the
quasi-circular hypothesis changes the primary spin posterior median
by only $\Delta a_1\simeq0.02$, much less than the corresponding
shift in $a_2$. The joint $m_1^{\rm src}$--$a_1$ constraints also
largely overlap between the two orbital hypotheses
(Fig.~\ref{fig:primary_spin}b).

These results reveal a strongly component-dependent response to the
orbital model. When eccentricity is excluded, the dominant change in
the component-spin magnitudes occurs in the secondary, whereas the
primary remains rapidly spinning. A natural interpretation is that
the primary is already concentrated near the high-spin boundary,
leaving comparatively little freedom for its spin magnitude to
accommodate additional non-circular waveform structure, while the
secondary spin can respond much more strongly.

The astrophysical implications of the two components are likewise
different. The eccentric secondary-spin posterior extends around
$a_2\simeq0.7$, close to the characteristic spin of a remnant from
a comparable-mass black hole merger with small progenitor spins.
Together with the secondary mass posterior overlapping the
pair-instability gap, as shown jointly with the spin constraint in Fig.~\ref{fig:spin}b, this provides a natural hierarchical merger
interpretation for the secondary.

The primary shows a different pattern. Its spin remains large under
both orbital hypotheses, and its mass posterior extends substantially
towards and beyond the upper part of the pair-instability-gap range.
It therefore does not exhibit the same remnant-like mass-spin
signature as the secondary. The two components may consequently
encode different formation histories, even if the present binary
was assembled dynamically.

\subsection*{Broader source parameter consistency}

The injection--recovery results described above show good agreement
with the GW231123 results not only for eccentricity and component
spins, but also across the broader source parameter space.
The corresponding one- and two-dimensional marginalized posteriors
are shown in Extended Data Figs.~\ref{fig:ed_ecc10}--\ref{fig:ed_qc55}.

These figures compare GW231123 with recoveries of the same
GW231123-like eccentric injection under the eccentric and
quasi-circular waveform hypotheses and across the two frequency
configurations considered in this work. In all cases, the recovered
source parameter constraints are consistent with the corresponding
GW231123 posteriors within their uncertainties.

This broader agreement provides an additional event-specific
consistency check, showing that the injected eccentric signal
reproduces not only the eccentricity and spin behaviour highlighted above, but also the overall source parameter constraints
inferred for GW231123.

Importantly, the injection--recovery analyses are performed in zero
noise. Their close agreement with the GW231123 results across the
11--448\,Hz and 20--448\,Hz eccentric analyses and the 11--448\,Hz
quasi-circular analysis shows that the injected signal closely
reproduces the real data inference. Since the injections contain no glitches, this consistency shows that the observed frequency- and waveform-dependent behaviour can be naturally reproduced by a genuinely eccentric GW231123-like signal without requiring a particular detector noise realization. Together with the LVK validation of the glitch-subtracted H1 strain, this provides an additional consistency check on the eccentric interpretation.

\section*{Discussion}

Previous studies have explored the possibility that GW231123 is
affected by point-mass lensing, although the lensing interpretation
remains uncertain \cite{LIGOScientific:2025cwb, Goyal:2025eqo, Chan:2025kyu, Shan:2025dcd, Cheung:2026pky}.
To examine whether our eccentricity inference remains stable when this
additional propagation effect is allowed, we extend the eccentric and
quasi-circular spin-precessing source models to include the wave-optics
amplification of an isolated point-mass lens.
The lens parameters are sampled jointly with the binary parameters,
using the same priors as in the LVK lensing analysis
\cite{LIGOScientific:2025cwb}, while retaining the detector calibration treatment adopted in the corresponding unlensed analysis (Methods).

With this extension, the eccentric model gives
$e_{5.5\,\mathrm{Hz}}=0.32^{+0.09}_{-0.10}$, consistent with the
corresponding unlensed result,
$e_{5.5\,\mathrm{Hz}}=0.34^{+0.12}_{-0.23}$.
The  log Bayes factor between the eccentric and quasi-circular hypotheses
when point-mass lensing is included is $\ln\mathcal{B}_{\rm ecc/qc}=2.27$, also consistent with that obtained for the corresponding unlensed models, showing that the eccentric
interpretation remains favoured when lensing is allowed.
At the same time, the posterior distributions of the redshifted lens mass and dimensionless source position  remain broad and largely prior dominated, indicating that the data provide little information on the lens parameters themselves.
Thus, allowing for point-mass lensing neither materially changes the inferred eccentricity nor the relative preference for eccentric over quasi-circular orbital dynamics.
More broadly, the LVK lensing analysis found that the support for
point-mass lensing varies substantially across waveform models \cite{LIGOScientific:2025cwb},
suggesting that part of the apparent lensing preference may reflect
source physics not captured by the waveform models used in those
analyses, potentially including orbital eccentricity.

The eccentric interpretation also changes the formation picture
suggested for GW231123. Previous population-informed studies based on
quasi-circular parameter estimation have interpreted both components
as products of extended hierarchical growth \cite{Li:2025fnf}.
Our results instead motivate a simpler, asymmetric picture.
Using the lower edge of the pair-instability gap inferred without
GW231123 itself, we estimate the maximum mass of
a simple second-generation remnant formed from two first-generation
black holes below this boundary as
$m_{\rm 2G,max}\simeq2m_{\rm low}(1-\epsilon_{\rm GW})$,
where $\epsilon_{\rm GW}$ is the fraction of the progenitor mass
radiated in gravitational waves.
We marginalize over the GW231123-excluded lower-edge posterior of
Tong et al.~\cite{Tong:2025wpz},
$m_{\rm low}=43.44^{+7.06}_{-5.53}\,M_\odot$, and adopt
$\epsilon_{\rm GW}=0.05$.
For each orbital hypothesis, we quantify the compatibility with this
simple $1{\rm G}+1{\rm G}\rightarrow2{\rm G}$ mass budget by the
fraction of the component-mass posterior satisfying
$m_i^{\rm src}<m_{\rm 2G,max}$.
For the secondary, the ratio of this fraction between the eccentric
and quasi-circular interpretations is about $2.2$, whereas for the
primary the compatibility remains only ${\cal O}(10^{-3})$.
Together with the shift of the secondary spin towards $a_2\simeq0.7$,
this makes a simple second-generation origin easier to accommodate
for the secondary, while the primary requires either formation beyond
the gap or a more complex hierarchical history.

This result directly affects how GW231123 enters population
measurements of the pair-instability mass gap.
Tong et al.~\cite{Tong:2025wpz} infer the gap from the
secondary-mass distribution and show that GW231123 is essential for
constraining its upper edge: removing the event leaves the lower edge
largely unchanged but makes the upper edge poorly constrained.
Our analysis indicates that the secondary of GW231123 may itself be a
hierarchical-merger remnant rather than a black hole formed directly
above the pair-instability gap.
If so, the secondary mass of GW231123 cannot by itself establish
the upper edge of the pair-instability mass gap.
More generally, when a population boundary is
constrained mainly by a few unusual events, waveform assumptions for those events can directly affect the interpretation of that boundary.

\section*{Methods}

\subsection*{Detector data and analysis band}

We analyse strain data from the LIGO Hanford (H1) and Livingston
(L1) detectors for GW231123
\cite{LIGOScientific:2025rsn,GWOSC}.
For each detector, we use an 8s data segment extending from
6\,s before to 2\,s after the trigger, sampled at 1,024\,Hz.

For H1, we use the publicly released BayesWave glitch-subtracted strain data and evaluate the likelihood over 11--448\,Hz. For L1, we use the original strain data and adopt a lower frequency cutoff of 20\,Hz, following the data-quality assessment of GW231123
\cite{LIGOScientific:2025rsn}.
This conservative choice excludes the low-frequency region affected by the reported 10–20 Hz L1 glitch, despite the LVK finding that the glitch had no measurable effect on parameter estimation, while retaining the
lower-frequency information in the glitch-subtracted H1 data. 
The event time satisfies the relevant O4a data-quality requirements.
In addition, the LVK-released H1 glitch-subtracted strain passed the
official validation comparing the residual noise with Gaussian noise,
with no evidence that the remaining broadband non-stationarity affected
the GW231123 analysis \cite{LIGOScientific:2025rsn}.
The use of different lower frequency cutoffs for different detectors has precedent in gravitational-wave parameter estimation. For example, different cutoffs were adopted for Hanford and Livingston in the analysis of GW170608 \cite{LIGOScientific:2017vox,De:2018zrk}.
Throughout this work, we use ``11--448\,Hz'' analysis as shorthand for the detector-specific configuration in which the likelihood is evaluated over 11--448\,Hz for H1 and 20--448\,Hz for L1.
As a robustness check, we also extend the L1 likelihood down to
11\,Hz, so that both detectors are analysed over 11--448\,Hz,
and obtain nearly the same eccentricity constraints.
Nevertheless, we present our main results obtained in the former case.

\subsection*{Waveform hypotheses}

We generate waveforms using the spin-precessing TEOBResumS--Dal\'i framework, including all $\ell\leq4$ gravitational wave modes provided by the model \cite{Nagar:2024dzj,Gamba:2024cvy,Jan:2025zcm}.
The eccentric and quasi-circular hypotheses use the same waveform family and identical settings for all shared parameters. In the quasi-circular hypothesis, the eccentricity is fixed to
zero, whereas in the eccentric hypothesis the orbital eccentricity and mean anomaly are sampled. Both hypotheses include spin
precession and the same higher-order modes.

We specify the orbital eccentricity at a reference $(2,2)$-mode gravitational wave frequency of 5.5\,Hz for the 11--448\,Hz analysis and 10\,Hz for the 20--448\,Hz analyses, denoting the corresponding
eccentricities by $e_{5.5\,\mathrm{Hz}}$ and $e_{10\,\mathrm{Hz}}$,
respectively. Waveform generation starts at these reference frequencies, corresponding to half the lowest likelihood frequency of each analysis configuration, to retain the in-band contributions of the included higher order modes.

\subsection*{Bayesian inference}

For detector data $d$ and waveform hypothesis $H$, the posterior is
\begin{equation}
p(\boldsymbol{\theta}\mid d,H)
=
\frac{
\mathcal{L}(d\mid\boldsymbol{\theta},H)
\,\pi(\boldsymbol{\theta}\mid H)
}{
Z_H
},
\end{equation}
where $\pi(\boldsymbol{\theta}\mid H)$ is the prior and
\begin{equation}
Z_H
=
\int
\mathcal{L}(d\mid\boldsymbol{\theta},H)
\,\pi(\boldsymbol{\theta}\mid H)
\,\mathrm{d}\boldsymbol{\theta}
\end{equation}
is the Bayesian evidence. Assuming stationary Gaussian noise that is independent between detectors, the log likelihood is
\begin{equation}
\ln\mathcal{L}
=
-\frac{1}{2}
\sum_k
\left\langle
d_k-h_k(\boldsymbol{\theta})
\mid
d_k-h_k(\boldsymbol{\theta})
\right\rangle_k
+\mathrm{const.},
\end{equation}
with the detector-specific noise-weighted inner product
\begin{equation}
\langle a\mid b\rangle_k
=
4\,\mathrm{Re}
\int_{f_{{\rm low},k}}^{f_{{\rm high},k}}
\frac{
\tilde a^{*}(f)\tilde b(f)
}{
S_{n,k}(f)
}
\,\mathrm{d}f .
\end{equation}
Here $S_{n,k}(f)$ is the one-sided noise power spectral density (PSD) of detector $k$. The waveform $h_k$ includes the detector response and the corresponding calibration corrections when calibration uncertainty is modelled.

For our analysis, we estimate the PSD independently for each detector using a median Welch estimator applied to the corresponding off-source strain data.
The LVK analysis of GW231123 instead used the median of the BayesWave/BayesLine PSD posterior, although the publicly released PSDs
for this event are available only above 20\,Hz
\cite{LIGOScientific:2025rsn,GWOSC}.
As a robustness check, we independently construct BayesWave/BayesLine median PSDs extending down to 11\,Hz.
Over the common 20--448\,Hz range, these PSDs closely agree with the LVK-released PSDs. Repeating the parameter estimation with these PSDs yields eccentricity constraints consistent with those obtained using the off-source median-Welch PSDs. Unless stated otherwise, the results reported here use the off-source median-Welch PSDs.

The injection–recovery analyses are performed in zero noise using the detector-specific PSDs adopted in the corresponding analyses, thereby isolating the frequency-dependent information content of the signal from any particular detector-noise realization.

We perform the inference with \texttt{Bilby}
\cite{Ashton:2018jfp} and the \texttt{dynesty} nested sampler
\cite{Speagle:2019ivv}. The production runs use 2,000 live
points and acceptance-walk sampling with
\texttt{naccept}$=60$. We define

\begin{equation}
\ln \mathcal{B}_{\rm A/B}
=\ln Z_{\rm A}-\ln Z_{\rm B},
\end{equation}
so that positive values favour hypothesis $A$.

The eccentricity prior is uniform over $e\in[0,0.5]$, corresponding
to a constant prior density of 2, and the mean anomaly is uniform
over $[0,2\pi]$. The eccentric and quasi-circular hypotheses use
identical priors for all shared source parameters. We sample the
detector-frame chirp mass over
$\mathcal{M}^{\rm det}\in[70,220]\,M_\odot$
and the mass ratio
$q=m_2/m_1\in[0.2,1]$.

The dimensionless spin magnitudes are
$a_i=c|\boldsymbol{S}_i|/(Gm_i^2)$, with $i=1,2$ and
$m_1\geq m_2$. They are uniform over $[0,0.99]$, and the spin
directions are isotropic. The luminosity-distance prior assumes a
constant merger rate per unit comoving volume and source-frame
time, with $d_L\in[300,8000]\,\mathrm{Mpc}$ and the Planck15
cosmology \cite{Planck:2015fie}. The sky position and viewing
direction are isotropic. The orbital phase and spin azimuths are
uniform over $[0,2\pi]$, while the polarization angle is uniform
over $[0,\pi]$. Source-frame masses are obtained
from detector-frame masses through
\begin{equation}
m^{\rm src}
=
\frac{m^{\rm det}}{1+z},
\end{equation}
with the redshift inferred from the luminosity distance using the
adopted cosmology.

\subsection*{Calibration uncertainty}

For all real data analyses, we marginalize over
frequency-dependent detector-calibration uncertainty using
cubic-spline corrections to the amplitude and phase response of each detector. We use ten spline nodes for H1 and ten for L1, with Gaussian priors constructed from the calibration-uncertainty data released for the event
\cite{LIGOScientific:2025rsn,GWOSC}.
The eccentric and quasi-circular hypotheses use the same
calibration prescription, and the calibration parameters are
sampled jointly with the source parameters.

\subsection*{Point-mass microlensing extension}

To test the robustness of the eccentric inference to an additional propagation effect, we also consider wave-optics microlensing by an isolated point-mass lens. The unlensed frequency-domain waveform
$\tilde h_{\rm UL}(f)$ is multiplied by the corresponding
amplification factor \cite{Takahashi:2003ix},
\begin{equation}
\tilde h_{\rm L}(f)
=
F(w,y)\tilde h_{\rm UL}(f),
\end{equation}
where
\begin{equation}
w=\frac{8\pi G M_L^z f}{c^3},
\end{equation}
$M_L^z=M_L(1+z_L)$ is the redshifted lens mass and $y$ is the dimensionless source position in units of the Einstein radius. For positive frequencies, the amplification factor is
\begin{align}
F(w,y)={}&
\exp\left[
\frac{\pi w}{4}
+\frac{iw}{2}
\left\{
\ln\left(\frac{w}{2}\right)-2\phi_m(y)
\right\}
\right]
\nonumber\\
&\times
\Gamma\left(1-\frac{iw}{2}\right)
{}_1F_1\left(
\frac{iw}{2},1;\frac{iwy^2}{2}
\right),
\end{align}
where
\begin{equation}
x_m=\frac{y+\sqrt{y^2+4}}{2},
\qquad
\phi_m(y)=\frac{(x_m-y)^2}{2}-\ln x_m .
\end{equation}
Here $\Gamma$ and ${}_1F_1$ denote the gamma and confluent
hypergeometric functions, respectively.

The lens parameters $M_L^z$ and $y$ are sampled jointly with the binary source parameters. The reported source-parameter constraints from this analysis are therefore conditional on the point-mass microlensing hypothesis.

\section*{Acknowledgements}

T.Y. acknowledges support from the National Natural Science Foundation
of China under Grant No.~12575063.
R.-G.C. is supported by the National Natural Science Foundation of China Grant No. 12588101 and No. 12535002.
Some of the numerical calculations in this work were performed using
the supercomputing system at the Supercomputing Center of Wuhan University.

\section*{Author contributions}

B.W. and T.Y. jointly conceived the study and developed the central scientific idea. B.W. carried out the data analysis and numerical calculations, generated the figures, and wrote the initial draft of the manuscript. T.Y. supervised the project throughout, guided the analyses and interpretation of the results, and contributed to the writing and revision of the manuscript. Z.C., B.H., X.S. and R.-G.C. provided scientific comments and suggestions and contributed to the revision of the manuscript. All authors reviewed and commented on the manuscript.

\section*{Competing interests}

The authors declare no competing interests.

\bibliographystyle{unsrtnat}
\bibliography{references}

@article{LIGOScientific:2025rsn,
    author = "Abac, A. G. and others",
    collaboration = "LIGO Scientific, VIRGO, KAGRA",
    title = "{GW231123: A Binary Black Hole Merger with Total Mass 190{\textendash}265 $M_{\odot}$}",
    eprint = "2507.08219",
    archivePrefix = "arXiv",
    primaryClass = "astro-ph.HE",
    reportNumber = "DCC: P2500026-v6, DCC: P2500026-v8",
    doi = "10.3847/2041-8213/ae0c9c",
    journal = "Astrophys. J. Lett.",
    volume = "993",
    number = "1",
    pages = "L25",
    year = "2025"
}

@article{Heger:2001cd,
    author = "Heger, A. and Woosley, S. E.",
    title = "{The nucleosynthetic signature of population III}",
    eprint = "astro-ph/0107037",
    archivePrefix = "arXiv",
    doi = "10.1086/338487",
    journal = "Astrophys. J.",
    volume = "567",
    pages = "532--543",
    year = "2002"
}

@article{Woosley:2016hmi,
    author = "Woosley, S. E.",
    title = "{Pulsational Pair-Instability Supernovae}",
    eprint = "1608.08939",
    archivePrefix = "arXiv",
    primaryClass = "astro-ph.HE",
    doi = "10.3847/1538-4357/836/2/244",
    journal = "Astrophys. J.",
    volume = "836",
    number = "2",
    pages = "244",
    year = "2017"
}

@article{Tong:2025wpz,
    author = "Tong, Hui and others",
    title = "{Evidence of the pair-instability gap from black-hole masses}",
    eprint = "2509.04151",
    archivePrefix = "arXiv",
    primaryClass = "astro-ph.HE",
    doi = "10.1038/s41586-026-10359-0",
    journal = "Nature",
    volume = "652",
    number = "8111",
    pages = "874--877",
    year = "2026"
}

@article{Rodriguez:2017pec,
    author = "Rodriguez, Carl L. and Amaro-Seoane, Pau and Chatterjee, Sourav and Rasio, Frederic A.",
    title = "{Post-Newtonian Dynamics in Dense Star Clusters: Highly-Eccentric, Highly-Spinning, and Repeated Binary Black Hole Mergers}",
    eprint = "1712.04937",
    archivePrefix = "arXiv",
    primaryClass = "astro-ph.HE",
    doi = "10.1103/PhysRevLett.120.151101",
    journal = "Phys. Rev. Lett.",
    volume = "120",
    number = "15",
    pages = "151101",
    year = "2018"
}

@article{Gerosa:2021mno,
    author = "Gerosa, Davide and Fishbach, Maya",
    title = "{Hierarchical mergers of stellar-mass black holes and their gravitational-wave signatures}",
    eprint = "2105.03439",
    archivePrefix = "arXiv",
    primaryClass = "astro-ph.HE",
    doi = "10.1038/s41550-021-01398-w",
    journal = "Nature Astron.",
    volume = "5",
    number = "8",
    pages = "749--760",
    year = "2021"
}

@article{Bartos:2016dgn,
    author = "Bartos, Imre and Kocsis, Bence and Haiman, Zolt and M{\'a}rka, Szabolcs",
    title = "{Rapid and Bright Stellar-mass Binary Black Hole Mergers in Active Galactic Nuclei}",
    eprint = "1602.03831",
    archivePrefix = "arXiv",
    primaryClass = "astro-ph.HE",
    doi = "10.3847/1538-4357/835/2/165",
    journal = "Astrophys. J.",
    volume = "835",
    number = "2",
    pages = "165",
    year = "2017"
}

@article{Samsing:2020tda,
    author = "Samsing, J. and Bartos, I. and D'Orazio, D. J. and Haiman, Z. and Kocsis, B. and Leigh, N. W. C. and Liu, B. and Pessah, M. E. and Tagawa, H.",
    title = "{AGN as potential factories for eccentric black hole mergers}",
    eprint = "2010.09765",
    archivePrefix = "arXiv",
    primaryClass = "astro-ph.HE",
    doi = "10.1038/s41586-021-04333-1",
    journal = "Nature",
    volume = "603",
    number = "7900",
    pages = "237--240",
    year = "2022"
}

@article{Jan:2025zcm,
    author = "Jan, Aasim and Nicolella, Sophia and Shoemaker, Deirdre and O'Shaughnessy, Richard",
    title = "{Measuring eccentricity and addressing waveform systematics in GW231123}",
    eprint = "2512.20060",
    archivePrefix = "arXiv",
    primaryClass = "gr-qc",
    doi = "10.1103/ly9b-w75v",
    journal = "Phys. Rev. D",
    volume = "113",
    number = "8",
    pages = "084052",
    year = "2026"
}

@misc{Chandra:2026voe,
      title={Eccentricity in Disguise? Insights from {GW231123} and Numerically Simulated Binary Black Hole Merger Signals}, 
      author={Koustav Chandra and Johann Fernandes and Akshita Mittal and Gregorio Carullo},
      year={2026},
      eprint={2606.15150},
      archivePrefix={arXiv},
      primaryClass={gr-qc},
      url={https://arxiv.org/abs/2606.15150}, 
}

@article{Peters:1964zz,
    author = "Peters, P. C.",
    title = "{Gravitational Radiation and the Motion of Two Point Masses}",
    doi = "10.1103/PhysRev.136.B1224",
    journal = "Phys. Rev.",
    volume = "136",
    pages = "B1224--B1232",
    year = "1964"
}

@article{Samsing:2017xmd,
    author = "Samsing, Johan",
    title = "{Eccentric Black Hole Mergers Forming in Globular Clusters}",
    eprint = "1711.07452",
    archivePrefix = "arXiv",
    primaryClass = "astro-ph.HE",
    doi = "10.1103/PhysRevD.97.103014",
    journal = "Phys. Rev. D",
    volume = "97",
    number = "10",
    pages = "103014",
    year = "2018"
}

@article{Antonini:2013tea,
    author = "Antonini, Fabio and Murray, Norman and Mikkola, Seppo",
    title = "{Black hole triple dynamics: breakdown of the orbit average approximation and implications for gravitational wave detections}",
    eprint = "1308.3674",
    archivePrefix = "arXiv",
    primaryClass = "astro-ph.HE",
    doi = "10.1088/0004-637X/781/1/45",
    journal = "Astrophys. J.",
    volume = "781",
    pages = "45",
    year = "2014"
}

@article{Chattopadhyay:2023pil,
    author = "Chattopadhyay, Debatri and Stegmann, Jakob and Antonini, Fabio and Barber, Jordan and Romero-Shaw, Isobel M.",
    title = "{Double black hole mergers in nuclear star clusters: eccentricities, spins, masses, and the growth of massive seeds}",
    eprint = "2308.10884",
    archivePrefix = "arXiv",
    primaryClass = "astro-ph.HE",
    doi = "10.1093/mnras/stad3048",
    journal = "Mon. Not. Roy. Astron. Soc.",
    volume = "526",
    number = "4",
    pages = "4908--4928",
    year = "2023"
}

@article{Romero-Shaw:2022fbf,
    author = "Romero-Shaw, Isobel M. and Gerosa, Davide and Loutrel, Nicholas",
    title = "{Eccentricity or spin precession? Distinguishing subdominant effects in gravitational-wave data}",
    eprint = "2211.07528",
    archivePrefix = "arXiv",
    primaryClass = "astro-ph.HE",
    doi = "10.1093/mnras/stad031",
    journal = "Mon. Not. Roy. Astron. Soc.",
    volume = "519",
    number = "4",
    pages = "5352--5357",
    year = "2023"
}

@article{Tibrewal:2026jci,
    author = "Tibrewal, Snehal and Zimmerman, Aaron and Lange, Jacob and Shoemaker, Deirdre",
    title = "{Misinterpreting spin precession as orbital eccentricity in gravitational-wave signals}",
    eprint = "2601.02260",
    archivePrefix = "arXiv",
    primaryClass = "gr-qc",
    reportNumber = "UT-WI-37-2025, LIGO-P2500767",
    doi = "10.1103/fh7x-xm5z",
    journal = "Phys. Rev. D",
    volume = "114",
    number = "6",
    pages = "064046",
    year = "2026"
}

@article{Divyajyoti:2025cwq,
    author = "Divyajyoti and others",
    title = "{Biased parameter inference of eccentric, spin-precessing binary black holes}",
    eprint = "2510.04332",
    archivePrefix = "arXiv",
    primaryClass = "gr-qc",
    reportNumber = "LIGO-P2500606",
    doi = "10.1103/jdkc-65lz",
    journal = "Phys. Rev. D",
    volume = "113",
    number = "10",
    pages = "103022",
    year = "2026"
}

@misc{Gamboa:2026ybr,
      title={Enabling gravitational-wave astronomy with spin-precessing black holes on generic orbits}, 
      author={Aldo Gamboa and Lorenzo Pompili and Alessandra Buonanno and Luca Sebastiani and Raffi Enficiaud and Michael Boyle and Lawrence E. Kidder and Harald P. Pfeiffer and Antoni Ramos-Buades and Mark A. Scheel},
      year={2026},
      eprint={2609.01568},
      archivePrefix={arXiv},
      primaryClass={gr-qc},
      url={https://arxiv.org/abs/2609.01568}, 
}

@misc{LIGOScientific:2025cwb,
      title={{GWTC-4.0}: Searches for Gravitational-Wave Lensing Signatures}, 
      author={Abac, A. G. and others},
      year={2026},
      eprint={2512.16347},
      archivePrefix={arXiv},
      primaryClass={gr-qc},
      url={https://arxiv.org/abs/2512.16347}, 
}

@misc{GWOSC,
  author        = {{LIGO Scientific Collaboration} and {Virgo Collaboration} and {KAGRA Collaboration}},
  title         = {{GW231123: a Binary Black Hole Merger with Total Mass $190$--$265\,M_\odot$}},
  howpublished  = {Zenodo, version v3},
  year          = {2025},
  doi           = {10.5281/zenodo.17437902},
  url           = {https://doi.org/10.5281/zenodo.17437902}
}

@article{LIGOScientific:2017vox,
    author = "Abbott, B. P. and others",
    collaboration = "LIGO Scientific, Virgo",
    title = "{GW170608: Observation of a 19-solar-mass Binary Black Hole Coalescence}",
    eprint = "1711.05578",
    archivePrefix = "arXiv",
    primaryClass = "astro-ph.HE",
    reportNumber = "LIGO-DOCUMENT-P170608-V8",
    doi = "10.3847/2041-8213/aa9f0c",
    journal = "Astrophys. J. Lett.",
    volume = "851",
    pages = "L35",
    year = "2017"
}

@article{De:2018zrk,
    author = "De, Soumi and Biwer, Christopher M. and Capano, Collin D. and Nitz, Alexander H. and Brown, Duncan A.",
    title = "{Posterior samples of the parameters of binary black holes from Advanced LIGO, Virgo{\textquoteright}s second observing run}",
    eprint = "1811.09232",
    archivePrefix = "arXiv",
    primaryClass = "gr-qc",
    doi = "10.1038/s41597-019-0086-6",
    journal = "Sci. Data",
    volume = "6",
    number = "1",
    pages = "81",
    year = "2019"
}

@article{Nagar:2024dzj,
    author = "Nagar, Alessandro and Gamba, Rossella and Rettegno, Piero and Fantini, Veronica and Bernuzzi, Sebastiano",
    title = "{Effective-one-body waveform model for noncircularized, planar, coalescing black hole binaries: The importance of radiation reaction}",
    eprint = "2404.05288",
    archivePrefix = "arXiv",
    primaryClass = "gr-qc",
    doi = "10.1103/PhysRevD.110.084001",
    journal = "Phys. Rev. D",
    volume = "110",
    number = "8",
    pages = "084001",
    year = "2024"
}

@article{Gamba:2024cvy,
    author = "Gamba, Rossella and Chiaramello, Danilo and Neogi, Sayan",
    title = "{Toward efficient effective-one-body models for generic, nonplanar orbits}",
    eprint = "2404.15408",
    archivePrefix = "arXiv",
    primaryClass = "gr-qc",
    doi = "10.1103/PhysRevD.110.024031",
    journal = "Phys. Rev. D",
    volume = "110",
    number = "2",
    pages = "024031",
    year = "2024"
}

@article{Ashton:2018jfp,
    author = "Ashton, Gregory and others",
    title = "{BILBY: A user-friendly Bayesian inference library for gravitational-wave astronomy}",
    eprint = "1811.02042",
    archivePrefix = "arXiv",
    primaryClass = "astro-ph.IM",
    doi = "10.3847/1538-4365/ab06fc",
    journal = "Astrophys. J. Suppl.",
    volume = "241",
    number = "2",
    pages = "27",
    year = "2019"
}

@article{Speagle:2019ivv,
    author = "Speagle, Joshua S.",
    title = "{dynesty: a dynamic nested sampling package for estimating Bayesian posteriors and evidences}",
    eprint = "1904.02180",
    archivePrefix = "arXiv",
    primaryClass = "astro-ph.IM",
    doi = "10.1093/mnras/staa278",
    journal = "Mon. Not. Roy. Astron. Soc.",
    volume = "493",
    number = "3",
    pages = "3132--3158",
    year = "2020"
}

@article{Planck:2015fie,
    author = "Ade, P. A. R. and others",
    collaboration = "Planck",
    title = "{Planck 2015 results. XIII. Cosmological parameters}",
    eprint = "1502.01589",
    archivePrefix = "arXiv",
    primaryClass = "astro-ph.CO",
    doi = "10.1051/0004-6361/201525830",
    journal = "Astron. Astrophys.",
    volume = "594",
    pages = "A13",
    year = "2016"
}

@article{Takahashi:2003ix,
    author = "Takahashi, Ryuichi and Nakamura, Takashi",
    title = "{Wave effects in gravitational lensing of gravitational waves from chirping binaries}",
    eprint = "astro-ph/0305055",
    archivePrefix = "arXiv",
    doi = "10.1086/377430",
    journal = "Astrophys. J.",
    volume = "595",
    pages = "1039--1051",
    year = "2003"
}

@article{Divyajyoti:2023rht,
    author = "Divyajyoti and Kumar, Sumit and Tibrewal, Snehal and Romero-Shaw, Isobel M. and Mishra, Chandra Kant",
    title = "{Blind spots and biases: The dangers of ignoring eccentricity in gravitational-wave signals from binary black holes}",
    eprint = "2309.16638",
    archivePrefix = "arXiv",
    primaryClass = "gr-qc",
    doi = "10.1103/PhysRevD.109.043037",
    journal = "Phys. Rev. D",
    volume = "109",
    number = "4",
    pages = "043037",
    year = "2024"
}

@article{Li:2025fnf,
    author = "Li, Yin-Jie and Tang, Shao-Peng and Xue, Ling-Qin and Fan, Yi-Zhong",
    title = "{GW231123: Likely a Product of Successive Mergers from {\ensuremath{\sim}}10 Stellar-mass Black Holes}",
    eprint = "2507.17551",
    archivePrefix = "arXiv",
    primaryClass = "astro-ph.HE",
    doi = "10.3847/1538-4357/ae4102",
    journal = "Astrophys. J.",
    volume = "999",
    number = "1",
    pages = "127",
    year = "2026"
}

@article{Goyal:2025eqo,
    author = "Goyal, Srashti and Villarrubia-Rojo, Hector and Zumalacarregui, Miguel",
    title = "{Across the Universe: GW231123 as a Magnified and Diffracted Black Hole Merger}",
    eprint = "2512.17631",
    archivePrefix = "arXiv",
    primaryClass = "astro-ph.GA",
    doi = "10.3847/2041-8213/ae93b1",
    journal = "Astrophys. J. Lett.",
    volume = "1008",
    number = "1",
    pages = "L12",
    year = "2026"
}

@article{Chan:2025kyu,
    author = "Chan, Juno C. L. and Ezquiaga, Jose Mar{\'\i}a and Lo, Rico K. L. and Bowman, Joey and Maga{\~n}a Zertuche, Lorena and Vujeva, Luka",
    title = "{Discovering gravitational waveform distortions from lensing: A deep dive into GW231123}",
    eprint = "2512.16916",
    archivePrefix = "arXiv",
    primaryClass = "gr-qc",
    doi = "10.1103/5292-948q",
    journal = "Phys. Rev. D",
    volume = "114",
    number = "2",
    pages = "L021502",
    year = "2026"
}

@misc{Shan:2025dcd,
      title={How significant is the lensing interpretation of {GW231123}?}, 
      author={Xikai Shan and Huan Yang and Shude Mao},
      year={2026},
      eprint={2512.19118},
      archivePrefix={arXiv},
      primaryClass={astro-ph.GA},
      url={https://arxiv.org/abs/2512.19118}, 
}

@misc{Cheung:2026pky,
      title={The diffraction-lensing interpretation of {GW231123} with astrophysical priors}, 
      author={Mark Ho-Yeuk Cheung and Digvijay Wadekar and Matias Zaldarriaga and Tejaswi Venumadhav and Javier Roulet and Ajit Kumar Mehta},
      year={2026},
      eprint={2607.21834},
      archivePrefix={arXiv},
      primaryClass={astro-ph.CO},
      url={https://arxiv.org/abs/2607.21834}, 
}


\clearpage
\onecolumn

\begingroup

\setcounter{figure}{0}
\renewcommand{\thefigure}{\arabic{figure}}
\renewcommand{\theHfigure}{extendeddata.\arabic{figure}}

\captionsetup[figure]{
  name={Extended Data Fig.},
  labelsep=edbar,
  labelfont=bf,
  textfont=normalfont,
  font=small,
  format=plain,
  justification=justified,
  singlelinecheck=false,
  skip=8pt
}


\begin{figure}[p]
  \centering
  \includegraphics[
    width=\textwidth,
    height=0.68\textheight,
    keepaspectratio
  ]{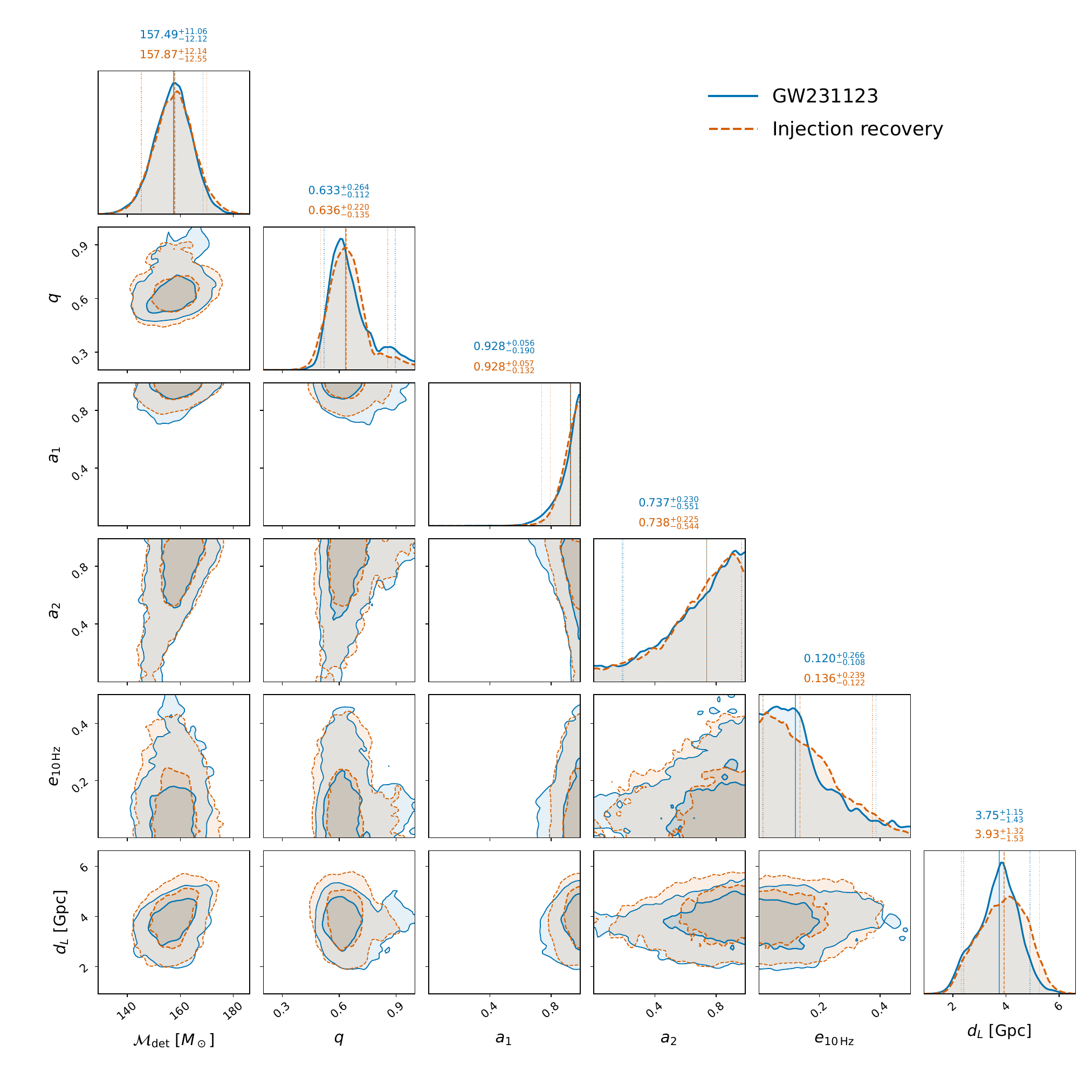}

  \caption{
\textbf{Source-parameter posteriors for GW231123 and the
GW231123-like eccentric injection in the 20--448\,Hz eccentric analyses.}
Blue solid curves show the GW231123 real-data posteriors, and orange
dashed curves show the recovery of the GW231123-like eccentric injection.
Both analyses use the eccentric waveform hypothesis over 20--448\,Hz,
with eccentricity defined at a 10\,Hz reference frequency.
The injection parameters are based on the joint maximum-likelihood
configuration of our 11--448\,Hz eccentric analysis.
The plotted eccentricity is the recovered $e_{10\,\mathrm{Hz}}$ at the
10\,Hz reference frequency.
The parameters shown are the detector-frame chirp mass
$\mathcal{M}_{\mathrm{det}}$, mass ratio $q=m_2/m_1\leq1$,
dimensionless primary- and secondary-spin magnitudes $a_1$ and $a_2$,
eccentricity $e_{10\,\mathrm{Hz}}$, and luminosity distance $d_L$.
Diagonal panels show one-dimensional marginalized posteriors.
Vertical solid and dashed lines mark the corresponding posterior medians,
dotted lines mark the 5th and 95th percentiles, and matching-colour
annotations give the medians and 90\% credible intervals.
In the off-diagonal panels, the inner and outer contours enclose the
50\% and 90\% highest-posterior-density credible regions, respectively.
}
  \label{fig:ed_ecc10}
\end{figure}

\clearpage


\begin{figure}[p]
  \centering
  \includegraphics[
    width=\textwidth,
    height=0.68\textheight,
    keepaspectratio
  ]{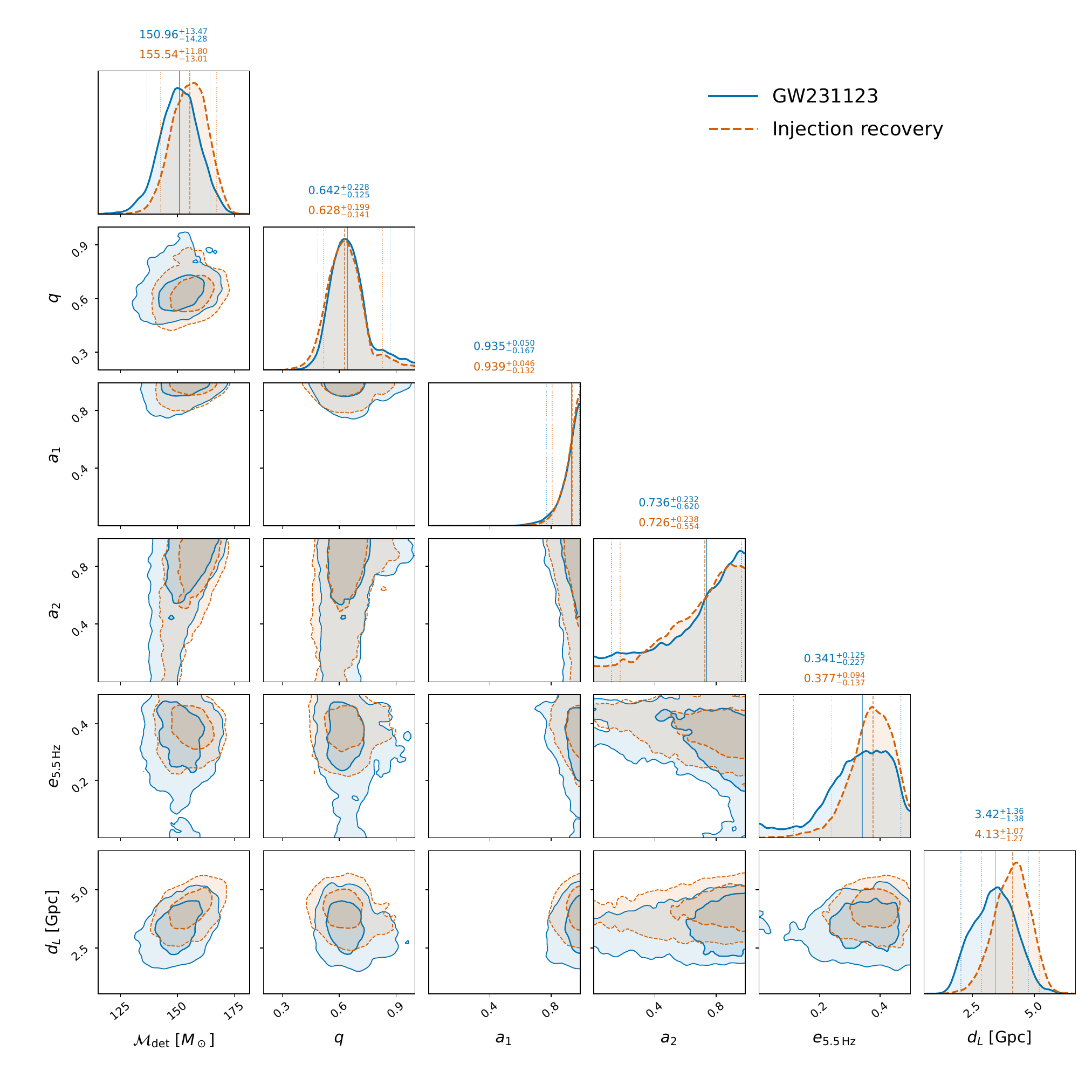}

 \caption{
\textbf{Source-parameter posteriors for GW231123 and the
GW231123-like eccentric injection in the 11--448\,Hz eccentric analyses.}
Blue solid curves show the GW231123 real-data posteriors, and orange
dashed curves show the recovery of the GW231123-like eccentric injection.
Both analyses use the eccentric waveform hypothesis in the
11--448\,Hz analysis configuration, with eccentricity defined at a
5.5\,Hz reference frequency.
The injection parameters are based on the joint maximum-likelihood
configuration of our 11--448\,Hz eccentric analysis.
The parameters shown are the detector-frame chirp mass
$\mathcal{M}_{\mathrm{det}}$, mass ratio $q=m_2/m_1\leq1$,
dimensionless primary- and secondary-spin magnitudes $a_1$ and $a_2$,
eccentricity $e_{5.5\,\mathrm{Hz}}$, and luminosity distance $d_L$.
Diagonal panels show one-dimensional marginalized posteriors.
Vertical solid and dashed lines mark the corresponding posterior medians,
dotted lines mark the 5th and 95th percentiles, and matching-colour
annotations give the medians and 90\% credible intervals.
In the off-diagonal panels, the inner and outer contours enclose the
50\% and 90\% highest-posterior-density credible regions, respectively.
}
  \label{fig:ed_ecc55}
\end{figure}

\clearpage


\begin{figure}[p]
  \centering
  \includegraphics[
    width=\textwidth,
    height=0.68\textheight,
    keepaspectratio
  ]{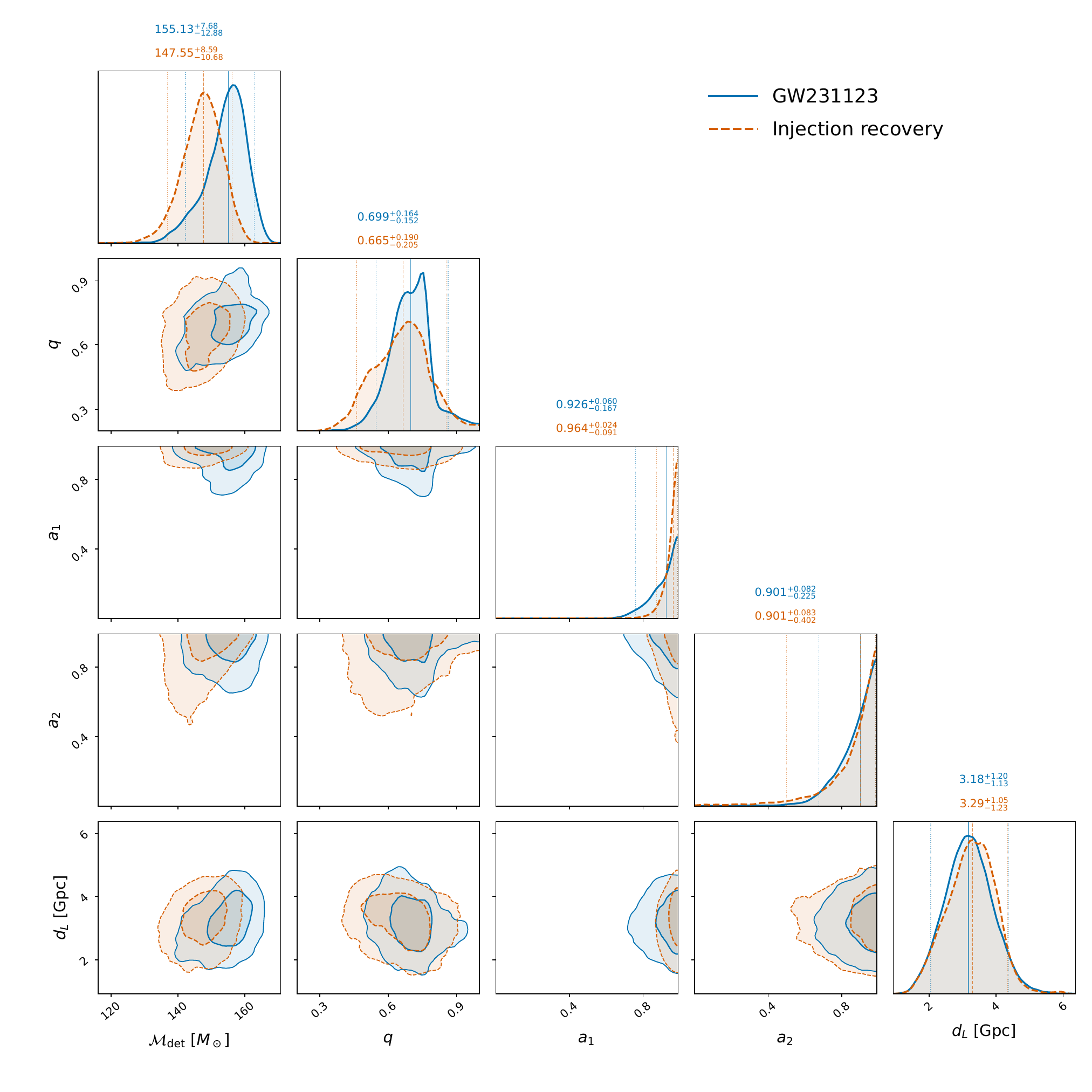}

  \caption{
\textbf{Source-parameter posteriors for GW231123 and the
GW231123-like eccentric injection under the quasi-circular hypothesis.}
Blue solid curves show the GW231123 real-data posteriors, and orange
dashed curves show the recovery of the GW231123-like eccentric injection.
Both analyses use the quasi-circular waveform hypothesis in the
11--448\,Hz analysis configuration.
The injection parameters are based on the joint maximum-likelihood
configuration of our 11--448\,Hz eccentric analysis, while eccentricity
is fixed to zero in the recovery model and is therefore not shown.
The parameters shown are the detector-frame chirp mass
$\mathcal{M}_{\mathrm{det}}$, mass ratio $q=m_2/m_1\leq1$,
dimensionless primary- and secondary-spin magnitudes $a_1$ and $a_2$,
and luminosity distance $d_L$.
Diagonal panels show one-dimensional marginalized posteriors.
Vertical solid and dashed lines mark the corresponding posterior medians,
dotted lines mark the 5th and 95th percentiles, and matching-colour
annotations give the medians and 90\% credible intervals.
In the off-diagonal panels, the inner and outer contours enclose the
50\% and 90\% highest-posterior-density credible regions, respectively.
}
  \label{fig:ed_qc55}
\end{figure}

\clearpage
\endgroup

\end{document}